\documentclass[a4paper,10pt,twoside]{cpc-hepnp}
\usepackage{multicol}
\usepackage{graphicx}
\usepackage{booktabs}
\usepackage{amssymb,bm,mathrsfs,bbm,amscd}
\usepackage[tbtags]{amsmath}
\usepackage{lastpage}

\begin{document}



\title{Mass differences and neutron pairing in Ca, Sn and Pb isotopes}

\author{%
B.\,S. Ishkhanov,$^{1, 2}$%
\quad S.\,V. Sidorov,$^{1}$%
\quad T.\,Yu. Tretyakova,$^{2;1)}$\email{tretyakova@sinp.msu.ru}%
\quad E.\,V. Vladimirova$^{2}$%
}
\maketitle

\address{%
{$^1$ 
Faculty of Physics, Lomonosov Moscow State University, Moscow 119991, Russia}\\
{$^2$Skobeltzyn Institute of Nuclear Physics,
Lomonosov Moscow State University, Moscow 119991, Russia}\\
}

\begin{abstract}
Various estimates of the even-odd effect of the mass shell of atomic nuclei are considered. Based on the experimental mass values of the Ca, Sn, and Pb isotopes, the dependence of the energy gap on the neutron number is traced and the relationship of this characteristic to the properties of external neutron subshells is shown. In nuclei with closed proton shells, effects directly related to neutron pairing and effects of nucleon shells are discussed.
\end{abstract}



\footnotetext[0]{\hspace*{-3mm}\raisebox{0.3ex}{$\scriptstyle\copyright$}2017 Chinese Physical Society and the Institute of High Energy Physics of the Chinese Academy of Sciences and the Institute of Modern Physics of the Chinese Academy of Sciences and IOP Publishing Ltd}%

\begin{multicols}{2}

\section{Introduction}

The creation of the shell model of the atomic nucleus {\citep {M49, J49}} is one of the most significant achievements of theoretical nuclear physics. The first attempts at its development were based on the model of atomic electron shells. The prospects for this approach were not that obvious, since there is a significant difference between the electrons in the atom and the nucleons in the atomic nucleus. In the case of an atom, the electrons are in the strong Coulomb field of the atomic nucleus, and the interactions of electrons with one another are a correction to the total potential (the ``screening" of the nuclear field by the electrons is very important). In the case of an 
atomic nucleus, the total self-consistent field is the result of nucleon-nucleon interactions and effectively takes its properties into account. Accordingly, the total atomic nucleus potential changes with transitions from isotope to isotope.

For a correct description of the properties of atomic nuclei, in addition to changing the mean-field potential it is also necessary to take into account the residual interaction. This, in spite of its small value, is crucial in determining the specific properties of the system of nucleons. In the first approximation, the so-called pairing 
forces are considered as the residual interaction --- an effective short-range interaction, which leads to 
an increase of the binding energy of a pair of nucleons when summation of their spins gives the total moment $J=0$. 
The pairing of identical nucleons makes it possible to explain many experimental facts, including 
the spin $J^P = 0^+$ of all even-even nuclei and the enhanced stability of even-even isotopes \cite{EG76, RS04, BM71}.

\section{Even-odd staggering and nucleon pairing}

The increasing of stability of even-even nuclei leads to the stratification of the mass surface on three components: 
one for even-even nuclei, one for odd-odd nuclei and one intermediate for nuclei with odd mass number $A$. A systematic study of the binding 
energies of a nucleus $B(A)$ shows that for even-even nuclei the following rule is fulfilled: 
\begin{equation}\label{EOS}
B(A)>\frac{1}{2}\left[B(A+1)+B(A-1)\right].
\end{equation}
The observed even-odd mass staggering (EOS) has been extensively explored in the literature  \citep{SD98, RB99, Do01, BH03, QW16}. The EOS effect is generally associated with the pairing gap $\Delta$, as suggested by BCS theory. To estimate its value  various more or less averaged equations are used: 
 three-, four-  or five-point \citep{BM71, JH84, MN88, MN92, BR00} formulas (so-called indicators):
\begin{equation}\label{three_point}
\Delta_n^{(3)}(N)=\frac{(-1)^N}{2}[S_n(N)-S_n(N+1)],
\end{equation}
\begin{equation}\label{four_point}
\Delta_n^{(4)}(N)=\frac{(-1)^N}{4}[-S_n(N+1)+2S_n(N)-S_n(N-1)],\\
\end{equation}
\begin{equation}\label{five_point}
\begin{array}{rl}
\Delta_n^{(5)}(N)&=\frac12[\Delta_n^{(4)}(N)+\Delta_n^{(4)}(N+1)]=\\
&=\frac{(-1)^N}{8}[3S_n(N+1)-3S_n(N)+\\
&+S_n(N-1)-S_n(N+2)],
\end{array}
\end{equation}
where $S_n(N) = B(N) - B(N-1)$  is the neutron separation energy of a nucleus $(N,Z)$. 
In the formulas (\ref{three_point} - \ref{five_point}) for neutron EOS, the proton number $Z$ is fixed.
Similar formulas (here and below) for protons can be obtained by fixing the neutron number  $N$ and replacing 
$N$ by $Z$.
It is seen from the formulas above  that the expression (\ref{four_point}) is also an averaging between $\Delta_n^{(3)}(N)$ 
 and $\Delta_n^{(3)}(N-1)$.

The relations (\ref{three_point}) and (\ref{four_point}) were originally obtained in order to get an analytic 
dependence of EOS on $A$ to introduce it as an additional pairing term to the semi-empirical Bethe-Weizs\"acker mass surface formula. 
From this point of view, the values of  $\Delta_n^{(3)}$ fluctuate much more strongly depending on $A$, but the 
result of their approximation differs slightly from the results for $\Delta_n^{(4)} $  \cite{BM71}. So, the four-point
formula (\ref{four_point}) became the basis for describing the EOS effect, and consequently for describing the pairing effect, 
for a long time. In some modern calculations even more smoothing formulas are used, taking into account  five 
\cite{MN88, MN92, Du01} or six experimental binding energies of isotopes \cite {JH84}. An increase of the number of isotopes 
 does not significantly affect the EOS calculation result, but the expansion of the range of experimental data 
in the region far from stability can lead to the usage of experimental data with significant errors. Modern mass formulae use more complicated pair approximations depending not only on power of $A$ but also on isospin relations \citep{RG10, WL13}. 
The relationship between different variants of the EOS estimation, as well as various variants of the $\Delta^{(4)}(A)$ approximation 
for protons and neutrons, are considered in Refs.~\cite {HBG02, FB09, IS14, QW16}.

Many studies are devoted to the evaluation of both the direct nucleon pairing contribution to the EOS    
and the contributions of other microscopic effects \cite{SD98, HJ13, Do01, CQ15, CQ15a, AA14}. It is shown \cite{Do01, CQ15} 
that the best estimation for identical nucleons pairing in the even $N$ nucleus is 
the three-point indicator (\ref{three_point}) for neighbor odd neutron number $\Delta_n^{(3)}(N+1)$. This conclusion corresponds to the direct 
determination of the two-neutron pairing $\Delta_{nn}(N)$ as the difference between the two-neutron  separation energy $S_{2n}(Z,N)$ from the even-even nucleus and the doubled   neutron separation energy $S_n(Z,N-1)$ from the neighboring odd nucleus $(N-1,Z)$ \cite{Preston}:
\begin{equation}\label{nn}
\begin{array}{rl}
\Delta_{nn}(N)&=S_{2n}(N) - 2S_n(N-1) =\\
&=S_n(N) - S_n(N-1) =\\
&= 2\Delta_n^{(3)}(N-1),
\end{array}
\end{equation}
where $S_{2n}(N) = B(N) - B(N-2)$. 
This definition considers the nucleus as a core with a pair of external ``valence"$\,$ nucleons, and does not take 
into account how the mean-field potential changes when ``valence" $\,$ nucleons are added or removed.
   
It is known that the $\Delta_{nn}(N)$ dependence for even-even nuclei is much smoother, and it produces an EOS estimate lower than that given by other formulas. Also, unlike the others, this characteristic is diminished for closed-shell nuclei, 
which corresponds to the common expectation that the pairing at shell closure should decrease in connection with 
the level density reduction. It can be expected that $\Delta_{nn}(N)$ includes the mean field contributions to 
the least extent, but it is apparently impossible to completely exclude their influence \cite{QW16, Do01}. In Ref.~\cite{FB09} it was  noted that, as  $\Delta_{nn}(N)$  includes the second differences of the binding energies, its value may be non-zero even without EOS.

\section{Seniority model}

With the pairing phenomena taken into account, the $A$-nucleon system Hamiltonian is:
\begin{equation}
\hat H = \hat H_0 +\hat H_{pair},
\end{equation}
where $\hat H_0$ is the intrinsic singe-particle Hamiltonian, determined by the nucleus mean-field, 
and the residual interaction corresponds to the monopole pairing:
\begin{equation} \label{Hp} 
\hat H_{pair}=-G\hat P^{\dag}\hat P.
\end{equation}
Here $G$ is the pairing strength parameter, and $\hat P^\dag$ and $\hat P$ denote the pair creation 
and annihilation operators. A rough experimental estimate gives $G_n=25/A$~MeV for neutrons and 
$G_p=17/A$~MeV for protons \cite{EG76}. 

The seniority model \cite{R43, Ta93} is one of the simplest models. It describes the filling of a shell with the total 
angular momentum $j$ over a closed core. Following Ref.~\cite{Do01}, let us consider $n$ nucleons moving in a $2\Omega$-fold degenerated
shell ($2\Omega = 2j+1$), described by the Hamiltonian in Eq.~(\ref{Hp}). The energy eigenvalues in the seniority model can be expressed 
in terms of nucleon number $n$ and seniority $\nu$ 
(the number of unpaired nucleons (quasiparticles)  in the configuration considered):
\begin{equation} \label{E} 
E(n,\nu)=-\frac{1}{4}G(n-\nu)(2\Omega-\nu-n+2).
\end{equation}
The nuclear ground state has seniority  $\nu=0$ for even nucleon number $n$ (all nucleons are paired), and  $\nu=1$
for odd number $n$.
The value of EOS according to the formula for the three-point indicator (\ref{three_point}) is:
\begin{equation} \label{sen1}
\Delta_{\tau}^{(3)}(n)=
\begin{cases}
\frac12 G\Omega+\frac12 G & \text{for even $n$,} \\
\frac12 G\Omega & \text{for odd $n$.}
\end{cases}
\end{equation}
The index $\tau = n, p$ denotes the nucleon type. Since this result depends only on whether the number of particles $n$ is even or odd and does not depend on  the absolute value of $n$, the average  four- and five-point indicators (\ref {four_point})
and (\ref {five_point}) in the seniority model are the same:
\begin{equation} \label{sen3}
\Delta_{\tau}^5(n)= \Delta_{\tau}^4(n) = \frac{1}{2} G \Omega + \frac{1}{4} G, \quad \text{for all $n$.}
\end{equation}

The pairing energy direct estimation $\Delta_ {\tau\tau}$ (\ref{nn}) for an even
neutron number will be smaller than the doubled three-point indicator $2\Delta_{\tau} ^ {(3)}$:
\begin{equation} \label{sen2}
\Delta_{\tau\tau}(N)=
\begin{cases}
 G \Omega & \text{for even $n$,} \\
 G \Omega + G & \text{for odd $n$.}
\end{cases}
\end{equation}
In this case the pairing value is equal to the doubled EOS effect $\Delta_{\tau\tau}(n) = 2\Delta_{\tau}^{(3)}(n-1)$ and does not depend on the absolute value of $n$.

\section{Nucleon separation energy}

In the simplest case of two neutrons pairing over the closed core,  the pairing energy $\Delta_ {nn}(N)$ (\ref {nn}) corresponds to the
doubled EOS effect $\Delta_n ^{(3)}(N-1)$. Hereafter we consider the corresponding doubled indicators:
\begin{eqnarray}\label{Dnn}
\Delta^{(3)}_{nn} (N) = 2\Delta^{(3)}_n(N),\\
\Delta^{(4)}_{nn} (N) = 2\Delta^{(4)}_n(N),\\
\Delta^{(5)}_{nn} (N) = 2\Delta^{(5)}_n(N).\label{Dnn1}
\end{eqnarray}

Since relations (\ref{Dnn} - \ref{Dnn1}) depend on the nucleon separation energies, let us consider the 
neutron separation energy as a function of the neutron number $N$ in isotopes $Z=$~Const. In Fig.~\ref {Sn_Ca}(a),
the measured neutron separation energy $S_n$ in Ca isotopes $(Z = 20)$ is plotted. The dependence $S_n$ shows a saw-tooth form, as a consequence of the pairing effect. 

\begin{center}
\includegraphics[width=7.2cm]{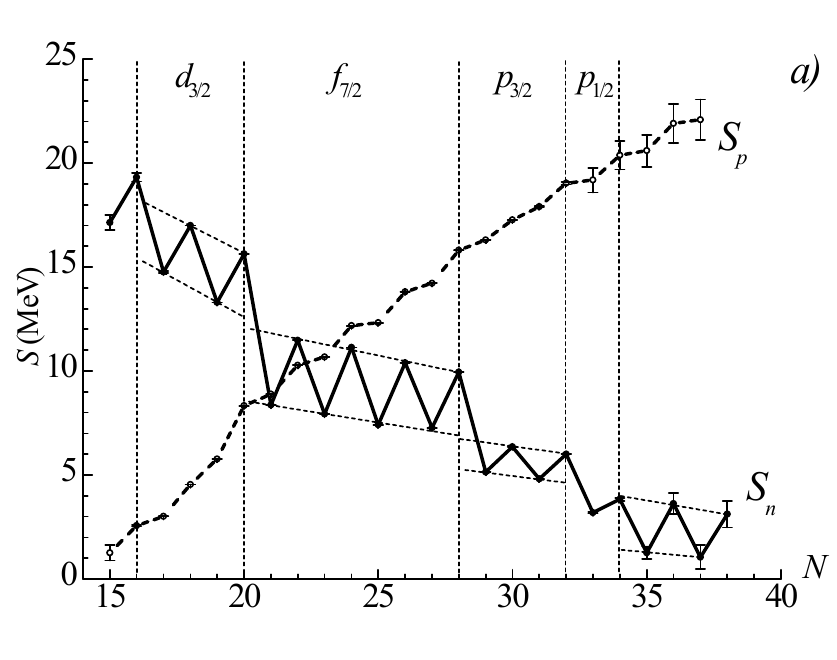}
\includegraphics[width=7.2cm]{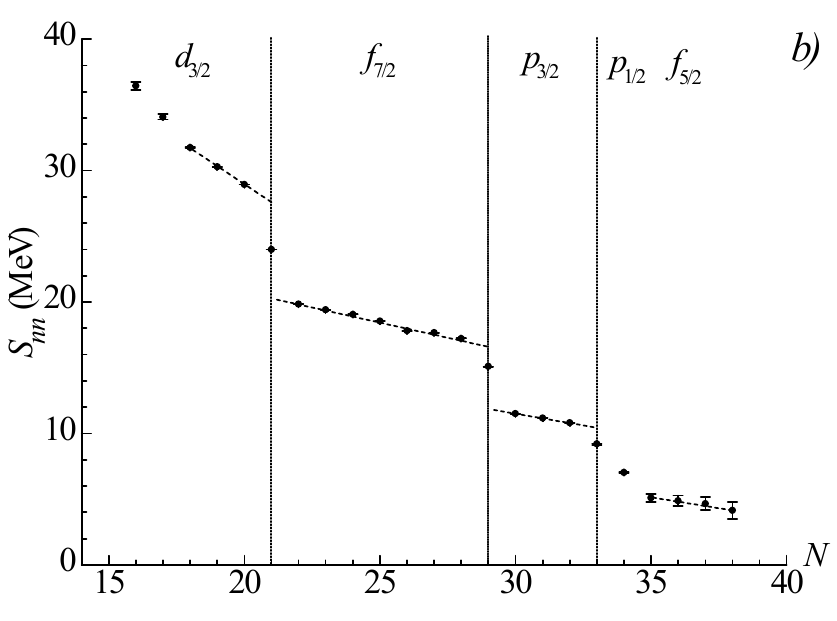}
\figcaption{Neutron $S_n(N)$, proton $S_p(N)$ (a) and two-neutron $S_{2n}(N)$ (b) separation energies
in Ca isotopes. Data from Ref.~\cite{AME16}. } \label{Sn_Ca}
\end{center}
The values  $S_n(N)$ for even and odd $N$ 
are divided into two groups lying well on two straight parallel lines. Sharp leaps between 
 groups of $S_n$ values for $ N =$ 20, 28, 32  correspond to subshell transitions. Since  
 the distance between single-particle levels is large in light nuclei, 
a consistent filling of the subshells $1d_{3/2} - 1f_{7/2} - 2p_{3/2}$ in Ca isotopes is traced well.

Fig.~\ref{Sn_Ca}(a) also gives the proton separation energy  $S_p(N)$. Despite the fact that the number of protons
remains constant at $ Z = 20 $, this dependence has a saw-tooth shape as well. Although it is not so pronounced
as that for $S_n(N)$, it nevertheless reflects the influence of the neutron pairing on the total mean-field potential changes.

The measured two-neutron separation energy $S_{2n}(N)$ (see Fig.~\ref{Sn_Ca}(b) does not show the effect of neutron pairing,
because  only  even $N$ or  odd $N$ isotopes are used for its calculation.

In Refs.~\cite{Ta56, Ta93} it was shown that in seniority model the energy of $n$ valence nucleons in the field of 
the closed core $B(j^n)$ can be expressed as 
\begin{equation}
\begin{array}{rl}
 B(j^n)= &B_{core}(n=0)+ n\varepsilon_j +\frac{n(n-1)}{2}\alpha-\\
 &-\frac{1}{2}\left[n-\frac{1-(-1)^n}{2}\right]\beta
\end{array}
\end{equation}
So the single nucleon separation energy 
\begin{equation}
\begin{array}{rl}
S_n(N) &= B(j^n)- B(j^{n-1})= \\
&=\varepsilon_j +(n-1)\alpha+\frac{1+(-1)^n}{2}\beta
\end{array}
\end{equation}
includes the energy $\varepsilon_j$, and depends on the kinetic energy of the nucleon on the $j$ shell and on the energy of interaction 
of an external nucleon with the core. The third term, proportional to $\beta$, corresponds to to the pairing effect, and 
the second one, proportional to $\alpha$, provides a common gradient of the curve $S_n(N)$. The values 
of the coefficients $\alpha$ and $\beta$ can be determined from the two-body matrix elements of ``valence'' nucleon interactions, so the pairing interaction not only determines the saw-tooth shape of 
$S_n(N)$,  but makes a contribution to the self-consistent mean field changes too. 
Sharp leaps between 
 groups of $S_n$ values for $ N =$ 20, 28, 32 are determined by the difference $\varepsilon_{j1} - \varepsilon_{j2}$ 
 in the transition between subshells $j_1$ and $j_2$.

\section{Identical nucleon pairing}

Due to the total gradient of the $S_n(N)$ dependence in one subshell, the pairing energy  $\Delta_{nn}$, obtained 
from (\ref{nn}), is always less than the result obtained from the three-point formula (\ref{three_point}) for even $N$:
$$ \Delta_{nn}<\Delta_{nn}^{(3)},$$ which is consistent with the seniority model (\ref{sen1}, \ref{sen2}).

In Fig.~\ref{Dn_Ca}, values of pairing energy indicators $\Delta_{nn}$ from (\ref{nn}), $\Delta_{nn}^{(3)}$ (\ref{three_point}) and 
$\Delta_{nn}^{(4)}$ (\ref{four_point}) in the Ca isotopes are plotted. All calculations were made on the base of measured nuclear masses from Ref.~\cite{AME16}. 
 If even and odd $N$ are considered together one can clearly see that $\Delta_{nn}$ and $\Delta_{nn}^{(3)}$ values coincide 
 accurarely on the $N=1$ shift (Fig.~\ref{Dn_Ca}(a)), and the  $\Delta^{(4)}_{nn}$ values are their average. 
The leap in $S_n(N)$ dependence due to the closure of the $1d2s$ subshell and the start of the $f_{7/2}$ subshell filling occurs at $N=20$ and $N+1 = 21$. As a result the three-point pairing energy indicator  $\Delta_{nn}^{(3)}$ (\ref{three_point}) has a sharp leap  even at $N=20$, but for  $\Delta_{nn}$ (\ref{nn}) the corresponding change is at odd $N+1=21$. That is why for even-even nuclei the three-point indicator $\Delta_{nn}^{(3)}$ has  significant fluctuations near the magic numbers, while the dependence of $\Delta_{nn}(N)$ has a smoother behavior (see Fig.~\ref{Dn_Ca}(b)). The values of the averaged indicators $\Delta^{(4)}_{nn}$ and $\Delta^{(5)}_{nn}$ are almost the same, but it should be noted that an increase in the number of points used to calculate the averaged characteristics narrows the range of isotopes under consideration.

\begin{center}
\includegraphics[width=7.2cm]{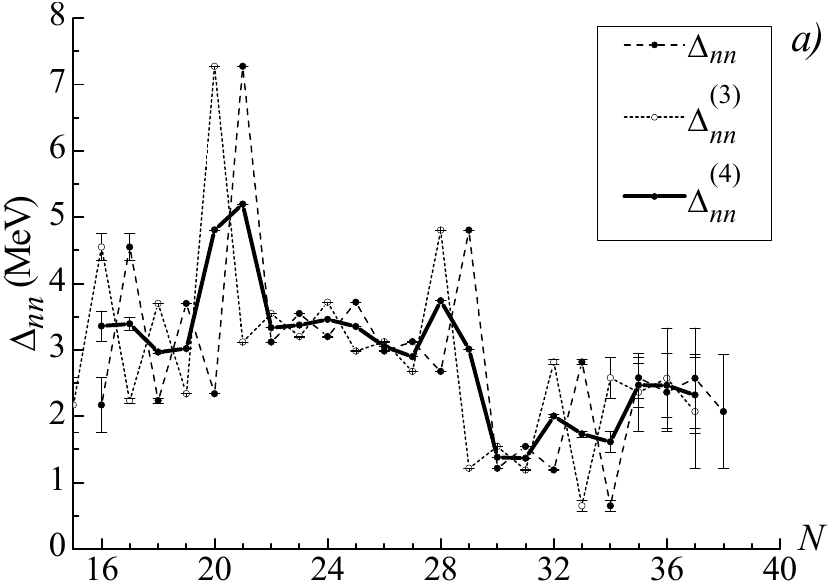}
\includegraphics[width=7.2cm]{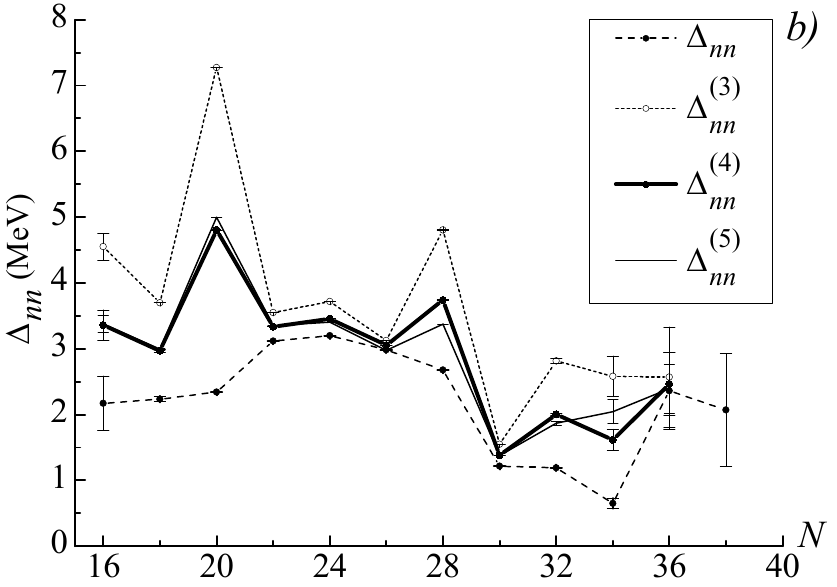}
\figcaption{Neutron pairing energy indicators $\Delta_{nn}^{(i)}$ (\ref{Dnn} - \ref{Dnn1}) in Ca isotopes 
for $N=16-38$ (a) and for even $N$ (b). Data 
from Ref.~\cite{AME16}.}\label{Dn_Ca}
\end{center}

It of interest is to consider the behavior of the difference between $\Delta_{nn}$ and $\Delta_{nn}^{(3)}$, formally coinciding with the pairing strength parameter $G$ in the simplest seniority model (\ref{sen1}):
 $$\delta e(N)= (-1)^N\left(\Delta_{nn}^{(3)}-\Delta_{nn} (N)\right).$$
At the same time, the definitions (\ref{three_point}, \ref{nn}) imply 
$$\delta e(N) = S_n(N-1)-S_n(N+1).$$
As the behavior of the $S_n(N)$ dependence (Fig.~\ref{Sn_Ca}(a)) shows, the value $\delta e(N)$  excludes the pairing effect and can be regarded as a correction associated with the core polarization and/or the contribution of the three-body interaction \cite{Br13}.

\section{Results for semimagic nuclei}

In Fig.~{\ref{Ca}} the dependencies $\Delta^{(3)}_{nn}$, $\Delta_{nn}$ and $\delta e$ in even-even Ca isotopes are plotted. In  Fig.~{\ref{Ca}}(a), in addition to $\Delta^{(3)}$, the experimental values of the first excited states $J^{\pi} =2^+$ are also given. In the  $^{40}$Ca case, which is typical for doubly magic nuclei, the $2^+$ state is not always the first excited state, which is associated with increasing of the rigidity and spherical symmetry of nuclei with filled shells \cite{BM71}. 

\begin{center}

\includegraphics[width=7.3cm]{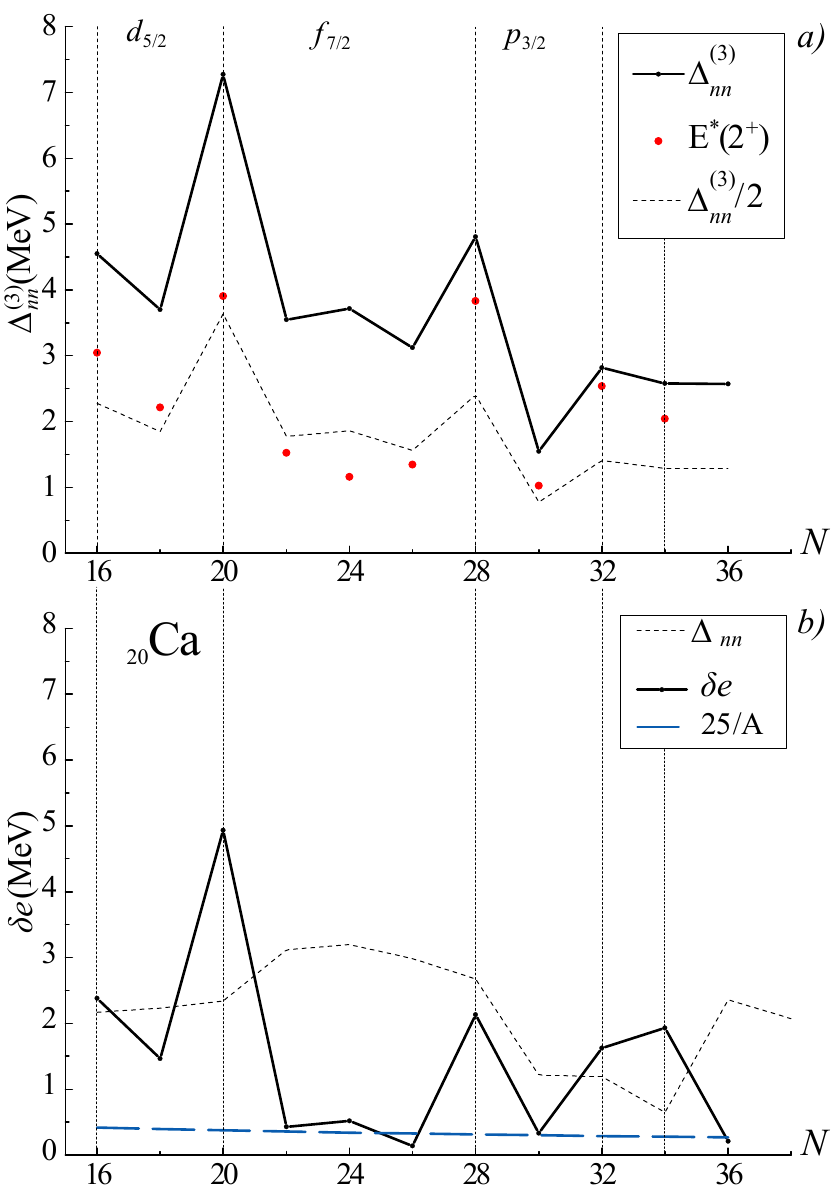}

\figcaption{Neutron pairing energies in Ca even-even isotopes. (a) The solid line corresponds to the three-point indicator $\Delta^{(3)}_{nn}$, and circles mark the experimental data $E_x(2_1^+)$\cite{ENSDF}, \cite{CDFE}. For comparison, the values $\Delta_n^{(3)}=\Delta^{(3)}_{nn}/2$ are indicated by the dashed line. 
(b)  The values $\Delta_{nn}$ (dashed line) and $\delta e$ (solid line) are plotted. For comparison, the values $\delta e(N)=25/A$ (MeV) are indicated by the dashed blue line.  }  
\label{Ca}
\end{center}

The spectroscopy of Ca isotopes  was considered in detail in Ref.~\cite{IS14a}. The low-energy spectra of odd Ca isotopes and the single-particle structure demonstrate the isolation of the subshell $f_{7/2}$ with respect to the closed core $^{40}_{20}$Ca, leading to a pronounced sequential filling of neutron subshells. In Fig.~{\ref{Ca}} the vertical dashed lines denoting the subshells filling correspond strictly to the maxima in the  $\Delta^{(3)}_{nn}$, $\delta e$ and $E_x({2^+})$ dependencies on the neutron number in the Ca isotopes. Thus, all three characteristics strongly correlate with each other. 

As mentioned above, the $\Delta_ {nn} (N) $ value for even nuclei has a more smoothed character, but, nevertheless, it is rather complicated and undergoes significant changes at the shell boundaries. One should note the similarity of values of $\Delta^{(3)}_{nn}$ and $\Delta_{nn}$ (and, correspondingly, small $\delta e$ value) for isotopes $^{42,44,46}$Ca.  An approximation which considers the closed core $^{40}_{20}$Ca with $f_{7/2}$ shell  filled consequently fits well for these isotopes \cite{MBZ64, EZ, II16} and one can assume that in this case the indicators $\Delta^{(3 )}_{nn}$ and $\Delta_{nn} $ (and respectively their averaging four- and five-point indicators $\Delta_{nn}^{(4)}$ and $\Delta_{nn}^{(5)}$)  reflect the neutron pairing effect most accurately. The values $\Delta_{nn}(N)$ more clearly demonstrate the dependence of the pairing energy on $j$ quantum number. The ratio between the values for different subshells corresponds to the ratio of the number of projections for the corresponding $j$ \cite{M50}:
$$\frac{\Delta_{nn}}{2j+1}=\frac{\Delta_{nn}(22)}{8}\approx
\frac{\Delta_{nn}(30)}{4}\approx\frac{\Delta_{nn}(36)}{6}\approx 0.35$$
A value of 0.35 agrees well with the accepted approximation of the neutron pairing strength parameter $ G_n\sim 25/A$~MeV (denoted in Fig.~\ref{Ca}, \ref{Sn} by a blue dashed line).

The behavior of indicators $\Delta^{(3)}_{nn}$ and $\delta e$ for semi-magic isotopes Sn and Pb with $Z = 50, 82$ have the same features. Figure~\ref {Sn} (a, b) shows the dependencies of $\Delta^{(3)}_{nn}$ and $E_x (2^+)$ on the neutron number $N$ in tin isotopes. For clarity, the dotted line also plots the value $ \Delta_n^{(3)} = \Delta^{(3)}_{nn}/2$, which  corresponds  with good accuracy to the excitation energy $ E_x (2 ^ +) $ for most isotopes in the chain.

\begin{center}
\includegraphics[width=7.2 cm]{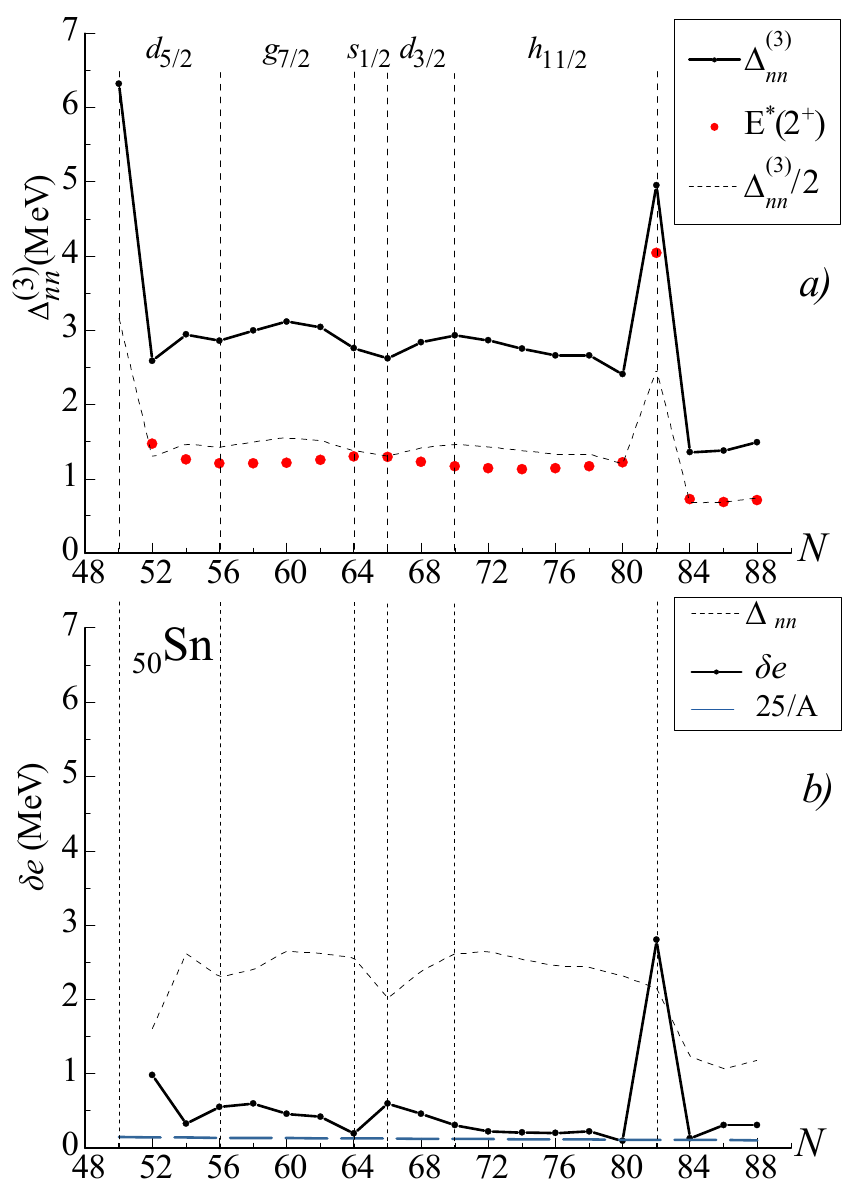}
\figcaption{Neutron pairing energies in  even-even  isotopes of tin Sn ($Z=50$).  For definitions see Fig.~\ref{Ca}.}
\label{Sn}
\end{center}

\begin{center}
\includegraphics[width=7.2 cm]{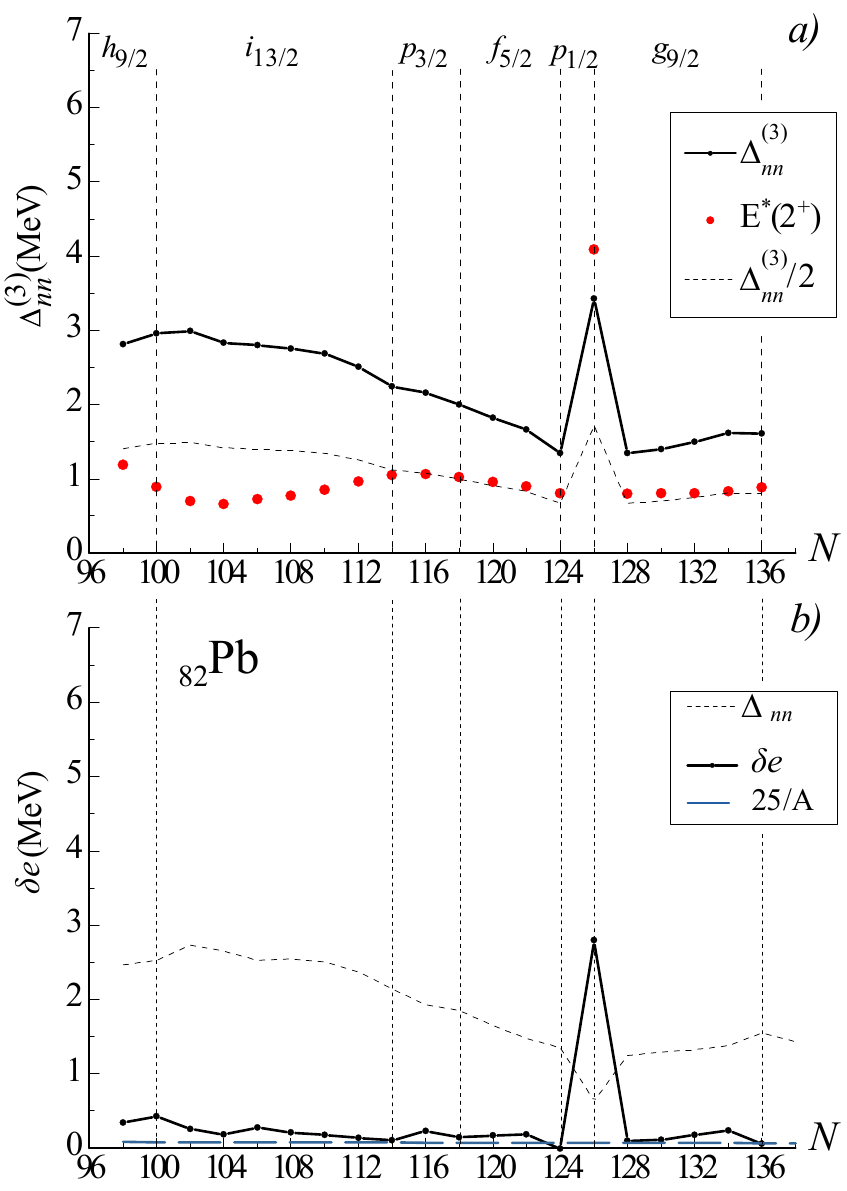}
\figcaption{Neutron pairing energies in  even-even  isotopes of lead Pb ($Z=82$).  For definitions see Fig.~\ref{Ca}.}
\label{Pb}
\end{center}

From the single-particle structure point of view,  the consequential filling of subshells does not exist in tin isotopes. Occupations of single-particle orbitals rise rather smoothly   with neutron number from $N=50$ to $N=80$ \cite {Is13}. Consequently, the $S_n(N)$ and $S_{2n}(N)$ dependencies have a smoothed variation without significant gaps associated with transitions between the subshells. In the $\Delta^{(3)}_{nn}(N)$ dependence, there is a small leap at $N = 66$, indicating the presence of a gap between the ($d_{5/2}, g_{7/2}$) and ($ s_{1/2}, d_{3/2}, h_{11/2} $) subshell groups. One should also note the proximity of the values and the explicit correspondence of the form of  $\Delta^ {(3)}_{nn} $ and $\Delta_ {nn}$ dependencies throughout the shell.  Nevertheless, $\delta e$ values  have pronounced changes, but they are minimal for $N>70$. In this region, the subshells with large values of $j$, $1h_{11/2} $ and $ 2f_{7/2} $, are filled, which leads to the characteristic spectra of the low-energy excited states  in these isotopes \cite {II16a}. A sharp leap in the three-point indicator $ \Delta^{(3)}_{nn}(N)$ for $N=82$ values corresponds to the transition to a new shell, and the decrease in the pairing effect that occurs can be related to a decrease in the number of projections $j$ on the outer shells \cite {Br15}, 16 on the subshell ($d_{3/2}, h_{11/2}$) compared to 8 on the more isolated subshell $f_{7/2}$:
   $$\frac{\Delta_{nn}(76)}{16}\approx\frac{\Delta_{nn}(84)}{8}\approx 0.15$$
   
   The same regularities can be traced in the $\Delta^{(3)}_{nn}$, $\Delta_{nn}$ and $\delta e$ dependencies  for lead isotopes (Fig.~\ref {Pb} a, b). The behavior of $\Delta^{(3)}_{nn}$ and $\Delta_{nn}$ is almost the same, which leads to $\delta e \approx$~Const for most Pb isotopes. A general decrease in the pairing effect $\Delta_{nn}$ can be associated with the transition from filling the high-momentum subshell group ($i_{11/2}, p_ {3/2} $) to states with a smaller value of $j$, up to $j =1/2$ for $N = 124$. Of course the behavior of the characteristics under consideration for neutron-rich isotopes with $ N> 132 $ is very interesting. For example, in Ref.~\cite {Br13} it was shown that negative values of $\delta e$ may indicate a sharp change in the type of deformation of the nucleus during the transition from one isotope to another. However, the error in determining the neutron separation energies for these isotopes amounts to tens of percentages and it is somewhat premature to make unambiguous conclusions about the magnitude of the characteristics based on the difference of separation energies $S_n$.
   
 \section*{Summary}
\mbox{}\vspace{-\baselineskip}

The main features of various atomic nucleus characteristics based on the mass differences, the neutron separation energy and various options for calculating the mass-surface EOS effect have been considered in this paper. For semi-magic isotopes with $Z=20, 50$, and $82$, for example, the complex nature of the even-odd effect, which includes both the nucleon pairing and other mean-field effects such as shell and subshell filling or symmetry effects, has been shown. The behavior of the characteristics involving the neutron separation energies from two neighboring isotopes, $\Delta^{(3)}_{nn}$ and $\Delta_{nn}$, strongly depends on the properties of the external nucleons and reflects not only the nucleon correlations in the middle of the shell filling, but also the  closed shells and subshell formation as the nucleon number goes through the magic numbers.

The $\Delta_{nn} $ value for even-even nuclei has a smooth $N$ dependence, since it involves isotopes with the numbers $N$ and $N-1$ and in the case of even $N$ does not include the leap associated with a change in the neutron single-particle energy upon transition to the next subshell. The systematic underestimation of the EOS value calculated by the formula $\Delta_{nn} $, compared with other three-, four- and five point indicators ($ \Delta^{(3)}_{nn} $, $\Delta^{(4)}_{nn} $, $ \Delta^{(5)}_{nn}$) is in accordance with the conclusions of the simplest seniority model. The smallest discrepancy between the various variants of the calculation is observed in the middle of the subshell filling. In this case  the EOS value corresponds most closely to the pairing energy $ \Delta \approx G \Omega$, and the difference between the $ \Delta ^ {(3)}_{nn} - \Delta_{nn}$ corresponds to the pairing strength parameter $ G=25/A $. In this area, far from magic numbers, pairing is most vividly manifested. A characteristic manifestation of the pairing effect is the low-lying $2^+$ states of collective nature, which form an energy gap 1-2 MeV between the ground and first exited state in even-even nucleus spectra.

 \acknowledgments{The authors would like to thank Dr. D. Lanskoy,  M.~Stepanov and L.~Imasheva for useful discussions and technical support.}

\end{multicols}

\vspace{-1mm}
\centerline{\rule{80mm}{0.1pt}}
\vspace{2mm}

\begin{multicols}{2}

\end{multicols}

\clearpage
\end{document}